\documentclass[11pt]{article}
\usepackage{graphics,epsfig,color}
\usepackage[]{graphicx}
\usepackage{latexsym}
\usepackage{amsmath, amsthm, amsfonts, amssymb}
\usepackage{verbatim}

\begin{document}

\newcommand{\ext}{=_E}
\newcommand{\Q}{$\mathfrak{Q}$}
\newcommand{\ita}{\textit}
\newcommand{\lra}{\leftrightarrow}

\newtheorem{theo}{Theorem}[section]
\newtheorem{definition}[theo]{Definition}
\newtheorem{lem}[theo]{Lemma}
\newtheorem{prop}[theo]{Proposition}
\newtheorem{coro}[theo]{Corollary}
\newtheorem{exam}[theo]{Example}
\newtheorem{rema}[theo]{Remark}
\newtheorem{example}[theo]{Example}
\newtheorem{principle}[theo]{Principle}
\newcommand{\ninv}{\mathord{\sim}} 
\newtheorem{axiom}[theo]{Axiom}

\title{Logic, Geometry And Probability Theory}

\author{{\sc Federico Holik}$^1$}

\maketitle

\begin{center}

\begin{small}
1 - Center Leo Apostel for Interdisciplinary Studies and, Department
of Mathematics, Brussels Free University Krijgskundestraat 33, 1160 Brussels, Belgium \\

\end{small}
\end{center}

\vspace{1cm}

\begin{abstract}
\noindent We discuss the relationship between logic, geometry and
probability theory under the light of a novel approach to quantum
probabilities which generalizes the method developed by R. T. Cox to
the quantum logical approach to physical theories.
\end{abstract}
\bigskip
\noindent

\begin{small}
\centerline{\em Key words: quantum logic-lattice theory-geometry of
quantum mechanics-quantum probability}
\end{small}

\bibliography{pom}

\begin{thebibliography}{10}

\bibitem{vN-Hilbert-Nordheim} J. Lacki, \textit{Arch. Hist. Exact Sci.}
\textbf{54} (2000) 279-318.

\bibitem{BvN} G. Birkhoff and J. von Neumann, \textit{Annals Math.} {\bf 37} (1936)
823-843.

\bibitem{RedeiHandbook} M. Redei ``The Birkhoff-von Neumann Concept of quantum Logic'', in Handbook of Quantum Logic and Quantum Structures, K. Engesser, D. M. Gabbay and D. Lehmann, eds., Elsevier (2009).

\bibitem{Gudder-StatisticalMethods} S. P. Gudder, \textit{Stochastic
Methods in Quantum Mechanics} North Holland, New York - Oxford
(1979).

\bibitem{Cantor1} G. Cantor, Contributions to the creation of transfinite set
theory, \emph{Annals of Mathematics}, Volume: \textbf{46}, Number:
\textbf{4} (1895) 481-512.

\bibitem{dallachiaragiuntinilibro} M. L. Dalla Chiara, R. Giuntini,
and R. Greechie, {\it Reasoning in Quantum Theory}, Kluwer Acad.
Pub., Dordrecht, (2004).

\bibitem{vNLetterToBirkhoff} J. von Neumann, Letter to G.Birkhoff, November
3, quoted in \cite{RedeiWhyVNDidNotLikedHilbert}.

\bibitem{RedeiWhyVNDidNotLikedHilbert} M. Redei, \textit{Studies in the
History and Philosophy of Modern Physics} \textbf{27}, (1996)
493-510.

\bibitem{vNAmsterdanTalk} M. Rédei and M. Stoeltzner (Eds.): \textit{John von Neumann and the Foundations of Quantum Physics},
Kluwer Academic Publishers, Dordrecht, Boston, London, (2001).

\bibitem{mikloredeilibro} M. R\'{e}dei, \textit{Quantum Logic in Algebraic
Approach}, Kluwer Academic Publishers, Dordrecht, (1998).

\bibitem{Redei-Summers2006} M. R\'{e}dei and S. Summers, \textit{Studies in History and Philosophy of Science Part B: Studies in
History and Philosophy of Modern Physics} Volume \textbf{38}, Issue
\textbf{2}, (2007) 390-417.

\bibitem{mackey-book} G. Mackey \textit{Mathematical foundations of quantum mechanics} New York:
W. A. Benjamin (1963).

\bibitem{KolmogorovProbability} Kolmogorov, A.N. Foundations of Probability Theory; Julius
Springer: Berlin, Germany, (1933).

\bibitem{CoxPaper} Cox, R.T. Probability, frequency, and reasonable expectation. \textit{Am.
J. Phys.} \textbf{14}, (1946) 1-13.

\bibitem{CoxLibro} Cox, R.T. \textit{The Algebra of Probable Inference}; The Johns Hopkins
Press: Baltimore, MD, USA, (1961).

\bibitem{Symmetry} P. Goyal and K. Knuth, \textit{Symmetry} \textbf{3} (2),
(2011) 171-206, doi:10.3390/sym3020171.

\bibitem{RingsOfOperatorsI} F. J. Murray and J. von Neumann, ``On rings
of operators", \textit{Ann. Of Math.}, \textbf{37}, (1936) 116-229.

\bibitem{RingsOfOperatorsII} F. J. Murray and J. von Neumann, ``On rings of
operators II", \textit{Trans. Amer. Math. Soc.}, \textbf{41} (1937),
208-248.

\bibitem{RingsOfOperatorsIII} J. von Neumann, ``On rings of operators III``, \textit{Ann.
Math.}, \textbf{41}, (1940), 94-161.

\bibitem{RingsOfOperatorsIV} F. J. Murray and J. von Neumann, ``On rings of operators
IV", Ann. Of Math., \textbf{44}, (1943) 716-808.s

\bibitem{Holik-Saenz-Plastino} F. Holik, A. Plastino and M.
S\'{a}enz, \textit{Annals Of Physics}, Volume \textbf{340}, Issue
\textbf{1}, 293-310, January (2014).

\bibitem{ProjectiveGeometries} I. Stubbe and B.V. Steirteghem \textit{Propositional systems, Hilbert lattices
and generalized Hilbert spaces}, in Handbook Of Quantum Logic
Quantum Structures: Quantum Structures, K. Engesser, D.M. Gabbay and
D. Lehmann, eds., (2007).

\bibitem{jauch} J. M. Jauch, {\it Foundations of Quantum Mechanics}, Addison-Wesley, Cambridge, (1968).

\bibitem{piron} C. Piron, {\it Foundations of Quantum Physics}, Addison-Wesley, Cambridge,
(1976).

\bibitem{kalm83} G. Kalmbach, {\it Orthomodular Lattices}, Academic
Press, San Diego, (1983).

\bibitem{vadar68} V. Varadarajan, {\it Geometry of Quantum Theory I}, van Nostrand, Princeton, (1968).

\bibitem{vadar70} V. Varadarajan, {\it Geometry of Quantum Theory
II}, van Nostrand, Princeton, (1970).

\bibitem{HandbookofQL} \textit{Handbook Of Quantum Logic And Quantum Structures}
(Quantum Logic), Edited by K. Engesser, D. M. Gabbay and D. Lehmann,
North-Holland (2009).

\bibitem{aertsdaub1} D. Aerts and I. Daubechies, \textit{Lett. Math. Phys.}
{\bf 3} (1979) 11-17.

\bibitem{aertsdaub2} D. Aerts and I. Daubechies, \textit{Lett. Math. Phys.}
{\bf 3} (1979) 19-27.

\bibitem{FR81} C. H. Randall and D. J. Foulis, in {\it Interpretation and Foundations
of Quantum Theory}, H. Neumann, ed. (Bibliographisches Institut,
Mannheim, 1981) pp. 21-28.

\bibitem{vN} J. von Neumann, {\it Mathematical Foundations of Quantum
Mechanics}, Princeton University Press, 12th. edition, Princeton,
(1996).

\bibitem{ReedSimon} M. Reed and B. Simon, \textit{Methods of modern
mathematical physics} I: Functional analysis, Academic Press, New
York-San Francisco-London (1972).

\bibitem{Soler-1995} M. Sol\`{e}r, \textit{Communications in Algebra}
\textbf{23}, (1995) 219-243.

\bibitem{Redei99} M. R\'{e}dei, \textit{The Mathematical Intelligencer},
\textbf{21} (4), (1999) 7-12.

\bibitem{aczel-book} Aczél, J., \textit{Lectures on Functional Equations and Their
Applications}, Academic Press, New York, (1966).

\bibitem{Gleason} A. Gleason, J. Math. Mech. \textbf{6}, (1957) 885-893.

\end{thebibliography}

\section{Introduction}

The formalism of quantum mechanics (QM) achieved its rigorous
formulation after a series of papers by von Neumann, Jordan, Hilbert
and Nordheim \cite{vN-Hilbert-Nordheim}. Its final form was
accomplished in the monumental work of von Neumann \cite{vN}. But an
interpretation of quantum mechanics is still lacking, despite the
efforts of many researchers during the years.

In the axiomatic approach of von Neumann projection operators play a
key role. The spectral decomposition theorem
\cite{ReedSimon,mikloredeilibro} allows to associate a projection
valued measure to any quantum observable represented by a self
adjoint operator \cite{vN,mikloredeilibro}. It turns out that the
set of projection operators can be endowed with a lattice structure;
more specifically, they form an orthomodular lattice \cite{kalm83}.

The subsequent developments turn the attention of von Neumann to the
theory of \emph{rings of operators}, better known as \emph{von
Neumann algebras} \cite{mikloredeilibro}. It was an attempt of
generalizing certain algebraic properties of Jordan algebras
\cite{vN-Hilbert-Nordheim}. But it turned out that the theory of von
Neumann algebras was strongly related to lattice theory: in a series
of papers, Murray and von Neumann provided a classification of
factors\footnote{von Neumann algebras whose center is formed by the
multiples of the identity operator}using orthomodular lattices
\cite{RingsOfOperatorsI,RingsOfOperatorsII,RingsOfOperatorsIII,RingsOfOperatorsIV}.
Time showed that all kinds of factors would find physical
applications, as is the case of type II factors in statistical
mechanics or type III factors in the rigorous axiomatic approach to
Quantum Field Theory (QFT) \cite{Redei-Summers2006,mikloredeilibro}.
It is important to remark that essentially all the information
needed to develop a physical theory out of these algebras is
contained in the logico-algebraic structure of their lattices of
projection operators \cite{Redei-Summers2006}.

On the other hand, lattice theory is deeply connected to geometry:
projective geometry can be described in terms of lattices and
related also to vector spaces \cite{ProjectiveGeometries}. As an
example, any vector space has associated a projective geometry and a
lattice of subspaces. In particular, projection operators of the
Hilbert spaces used in QM form a lattice and their sets of pure
states form projective geometries. But von Neumann was not only
interested in Hilbertian projection lattices; as the investigation
continued, he turned his attention to more general geometries,
namely, continuous geometries \cite{RedeiHandbook,Redei99}. That is,
the geometries associated to the type II$_{1}$ factors found in the
classification theory of Murray-von Neumann. As an example of the
exotic characteristics of the more general factors, type II$_{1}$
algebras are non-atomic and the type III contain no non-trivial
finite projections. In this way, it could be said that the
generalization of algebras studied by von Neumann points in the
direction of a rather radical generalization of geometry. Using the
words of von Neumann:

\begin{quotation}
``I would like to make a confession which may seem immoral: I do not
believe absolutely in Hilbert space any more. After all,
Hilbert-space (as far as quantum-mechanical things are concerned)
was obtained by generalizing Euclidean space, footing on the
principle of ``conserving the validity of all formal rules". This is
very clear, if you consider the axiomatic-geometric definition of
Hilbert-space, where one simply takes Weyl's axioms for a
unitary-Euclidean-space, drops the condition on the existence of a
finite linear basis, and replaces it by a minimum of topological
assumptions (completeness $+$ separability). Thus Hilbert-space is
the straightforward generalization of Euclidean space, if one
considers the vectors as the essential notions.

Now we begin to believe, that it is not the vectors which matter but
the lattice of all linear (closed) subspaces. [...]

But if we wish to generalize the lattice of all linear closed
subspaces from a Euclidean space to in infinitely many dimensions,
then one does not obtain Hilbert space, but that configuration,
which Murray and I called ``case II$_{1}$". (The lattice of all
linear closed subspaces of Hilbert-space is our ``case
I$_{\infty}$".) And this is chiefly due to the presence of the rule

$$a\leq c \longrightarrow a \vee (b \wedge c) = (a \vee b) \wedge
c \,\,\,[\mbox{modularity}!]$$

This ``formal rule" would be lost, by passing to Hilbert
space!"\cite{vNLetterToBirkhoff}
\end{quotation}

\noindent In this way, we see how deeply connected is the quantum
logical approach to physics developed by von Neumann to the
development of geometry. But what is the meaning of \emph{Logic} and
\emph{Geometry} in this context? In this short article, we will
explore a possible answer to this question and relate it to a
\emph{generalized probability theory} \cite{Redei-Summers2006}.

Later on, the quantum logical approach of Birkhoff and von Neumann
was developed further by other researchers, giving rise to
monumental foundational works (as examples see
\cite{jauch,piron,mackey-book,vadar68,vadar70,Gudder-StatisticalMethods}).
The problem of compound quantum systems in the QL approach was
studied first in \cite{aertsdaub1,aertsdaub2,FR81}. For complete
expositions of the QL approach see for example
\cite{HandbookofQL,dallachiaragiuntinilibro,kalm83}. It is very
important to remark that C. Piron showed that any propositional
system can be coordinatized in a generalized Hilbert space
\cite{piron}. A later result by Sol\`{e}r asserts that, under
reasonable conditions, it can only be a Hilbert space over the
fields of the real numbers, complex numbers or quaternions
\cite{Soler-1995}. In this way, an operational propositional system
with suitably chosen axioms can be represented in a generalized
Hilbert space.

Probability measures can be defined in general von Neumann algebras
\cite{Gudder-StatisticalMethods,Redei-Summers2006}. A generalized
non-kolmogorovian probability calculus can be developed including
Kolmogorovian probabilities as a particular case (i.e., when the
algebra is commutative) \cite{Redei-Summers2006}. Thus, the approach
developed by von Neumann and others leads to an interesting
connection between logic, geometry and probability theory.

In Section \ref{s:QuotationLarge} we discuss a problem posed by von
Neumann regarding the foundations of logic and probability theory.
We largely quote von Neumann because, on the one hand, we think that
these unpublished works are not too well known. On the other hand,
the content of the quotation is useful for our purposes: our aim is
to outline an interpretation of it. In Section
\ref{s:GeneralizedProbabilities} we review and discuss the main
features of quantum probabilities (and compare them to the classical
ones). In Section \ref{s:PaperAnnals} we discuss a novel approach to
generalized probability theory \cite{Holik-Saenz-Plastino}, and
then, in Section \ref{s:GLP} we discuss its consequences for the
problem posed by von Neumann providing the general features of an
interpretation.

An important final remark is in order before starting. While it is
certainly true that there exist many different interpretations of
the quantum formalism, and that there exists the problem of
metaphysical underdetermination (a point of view that we endorse),
in this work we will restrict to what was called the ``standard" (or
``Copenhaguen") interpretation. Of course, the so called ``standard"
interpretation, is not a consistent interpretation at all: it is a
collection of ideas regarding the quantum formalism and a discussion
between its founding fathers. While there are many points which can
be identified clearly (like the absence of determinism, and the
quantum jumps), it is not in general a consistent set of ideas. The
content of the present work goes in the direction of developing a
perspective which has the standard interpretation as a starting
point, and tries to provide more consistency to it, and we will not
discuss other interpretations. By no means is an aim of this work to
present a definitive point of view on the problem of the
interpretation of the quantum formalism in detriment of other
interpretations.

\section{A problem posed by von Neumann}\label{s:QuotationLarge}

\noindent As is well known, any boolean algebra can be represented
in a set theoretical framework (as subsets of a given set). With
regard to this relationship, von Neumann asserted that

\begin{quotation}
\noindent ``And one also has the parallelism that logics corresponds
to set theory and probability theory corresponds to measure theory
and that a given system of logics, so given a system of sets, if all
is right, you can introduce measures, you can introduce probability
and you can always do it in very many different ways."(unpublished
work reproduced in \cite{vNAmsterdanTalk}, pp. 244).
\end{quotation}

\noindent In this way, the connection between Logic, Set Theory, and
Probability Theory is clear (see also Sections
\ref{s:GeneralizedProbabilities} and \ref{s:GLP} of this work). What
does this means? The definition of Cantor of a set reads

\begin{quotation}
\noindent ``A set is a gathering together into a whole of definite,
distinct objects of our perception [Anschauung] or of our thought
---which are called elements of the set." \cite{Cantor1}
\end{quotation}

\noindent A set is a collection of \emph{objects}, and the internal
logic governing them is \emph{classical logic}. And of course, this
also applies to \emph{things in space}, because the last is just a
particular case of a set theoretical approach to collections of
objects: in the ultimate level the classical organization of
experience in an Euclidean space-time ---as well as in the curved
background of General Relativity--- is an expression of classical
logic.

\noindent But the things chance radically in the quantum formalism,
as von Neumann pointed out

\begin{quotation}
\noindent ``In the quantum mechanical machinery the situation is
quite different. Namely instead of the sets use the linear sub-sets
of a suitable space, say of a Hilbert space. The set theoretical
situation of logics is replaced by the machinery of projective
geometry, which is in itself quite simple.

However, all quantum mechanical probabilities are defined by inner
products of vectors. Essentially if a state of a system is given by
one vector, the transition probability in another state is the inner
product of the two which is the square of the cosine of the angle
between them. In other words, probability corresponds precisely to
introducing the angles geometrically. Furthermore, there is only one
way to introduce it. The more so because in the quantum mechanical
machinery the negation of a statement, so the negation of a
statement which is represented by a linear set of vectors,
corresponds to the orthogonal complement of this linear space."
(unpublished work reproduced in \cite{vNAmsterdanTalk}, pp. 244).
\end{quotation}

\noindent von Neumann continues

\begin{quotation}
\noindent ``And therefore, as soon as you have introduced into the
projective geometry the ordinary machinery of logics, you must have
introduced the concept of orthogonality. This actually is rigorously
true and any axiomatic elaboration of the subject bears it out. So
in order to have logics you need in this set a projective geometry
with a concept of orthogonality in it.

In order to have probability all you need is a concept of all
angles, I mean angles other than 90º. Now it is perfectly quite true
that in geometry, as soon as you can define the right angle, you can
define all angles. Another way to put it is that if you take the
case of an orthogonal space, those mappings of this space on itself,
which leave orthogonality intact, leave all the angles intact, in
other words, in those systems which can be used as models of the
logical background for quantum theory, it is true that as soon as
all the ordinary concepts of logics are fixed under some isomorphic
transformation, all of probability theory is already fixed."
(unpublished work reproduced in \cite{vNAmsterdanTalk}, pp. 244).
\end{quotation}

\noindent Now we ask: what is the meaning of the connection between
Geometry and Logic in the above quotations? It is clear that in
confronting with the empirical propositions of QM we are facing
essentially a \emph{Geometry}, which is at the same time a
\emph{Logic}. But this Geometry is not the geometry of classical
space-time. Quite on the contrary, \emph{is the geometrical form in
which quantum events are organized}. And of course, \emph{this
geometrical form has an internal logical structuration, which is the
quantum logic}.

\noindent It is important to remark that this logic does not
necessarily denies the classical logic that we use when we think.
The word \emph{logic} above refers to the \emph{organization of
experience (phenomena)}. But what is the connection of all this with
probability theory? von Neumann suggested a clue as follows

\begin{quotation}
\noindent ``This means, however, that one has a formal mechanism, in
which logics and probability theory arise simultaneously and are
derived simultaneously." (unpublished work reproduced in
\cite{vNAmsterdanTalk}, pp. 245).
\end{quotation}

\noindent In the rest of this work we will discuss the implications
of a novel derivation of QL using the algebraic properties of the
propositional lattice of QM \cite{Holik-Saenz-Plastino}.

\section{Generalized probability
theory}\label{s:GeneralizedProbabilities}

\subsection{Kolmogorov}\label{s:Kolmogorov}

\noindent In this Section we introduce classical probability theory
using the axioms of Kolmogorov \cite{KolmogorovProbability}. Given
an outcome set $\Omega$, consider a $\sigma$-algebra $\Sigma$ of
subsets of $\Omega$. Then, a probability measure will be given by a
function $\mu$ such that

\begin{subequations}\label{e:kolmogorovian}
\begin{equation}
\mu:\Sigma\rightarrow[0,1]
\end{equation}
\noindent which satisfies
\begin{equation}
\mu(\emptyset)=0
\end{equation}
\begin{equation}
\mu(A^{c})=1-\mu(A),
\end{equation}

\noindent where $(\ldots)^{c}$ means set-theoretical-complement and
for any pairwise disjoint denumerable family $\{A_{i}\}_{i\in I}$

\begin{equation}
\mu(\bigcup_{i\in I}A_{i})=\sum_{i}\mu(A_{i})
\end{equation}

\end{subequations}

\noindent The triad $(\Omega,\Sigma,\mu)$ is called a
\emph{probability space} (to which we refer as a \emph{Kolmogorovian
probability}). It is possible to show that if $(\Omega,\Sigma,\mu)$
is a Kolmogorovian probability space, all usual properties of
classical probability can be derived. Of particular importance for
this work is the \emph{inclusion-exclusion principle}

\begin{equation}\label{e:SumRule}
\mu(A\cup B)=\mu(A)+\mu(B)-\mu(A\cap B)
\end{equation}

\noindent which can be derived from \ref{e:kolmogorovian}. The
logical version of \eqref{e:SumRule} reads

\begin{equation}\label{e:SumRuleLogical}
\mu(A \vee B)=\mu(A)+\mu(B)-\mu(A \wedge B)
\end{equation}

\noindent due to the direct correspondence between the connectives
of classical logic (``$\vee$" and ``$\wedge$") and set theoretical
union and intersection. This is a clear expression of what is said
in our first quotation to von Neumann in Section
\ref{s:QuotationLarge}.

\subsection{Quantum probabilities}\label{s:probabilities}

Let $\mathcal{P}(\mathcal{H})$ be the orthomodular lattice of
projection operators in a separable Hilbert space. In order to
define quantum probabilities, the following axioms on a function $s$
must be \emph{postulated} \cite{mikloredeilibro}

\begin{subequations}\label{e:nonkolmogorov}
\begin{equation}
s:\mathcal{P}(\mathcal{H})\rightarrow [0;1]
\end{equation}
\noindent such that:
\begin{equation}\label{e:Qprobability1}
s(\textbf{0})=0 \,\, (\textbf{0}\,\, \mbox{is the null subspace}).
\end{equation}
\begin{equation}\label{e:Qprobability2}
s(P^{\bot})=1-s(P),\end{equation} \noindent and, for a denumerable
and pairwise orthogonal family of projections ${P_{j}}$
\begin{equation}\label{e:Qprobability3}
s(\sum_{j}P_{j})=\sum_{j}s(P_{j}).
\end{equation}
\end{subequations}

\noindent How do we know that these axioms capture all the desired
features of quantum probabilities? Gleason's theorem \cite{Gleason}
gives us the answer: if $dim(\mathcal{H})\geq 3$, for any measure
$s$ satisfying \eqref{e:nonkolmogorov} there exists a positive
Hermitian trace class operator (of trace one) $\rho_{s}$, such that

\begin{equation}\label{e:bornrule2}
s(P):=\mbox{tr}(\rho_{s} P)
\end{equation}

\noindent And also the converse is true; using Eqn.
\eqref{e:bornrule2}, any positive trace class Hermitian operator of
trace one defines a measure satisfying \eqref{e:nonkolmogorov}.

\noindent A generalized probability calculus can be extended to all
orthomodular lattices \cite{Redei-Summers2006} and even to
$\sigma$-orthocomplemented orthomodular posets
\cite{Gudder-StatisticalMethods} in an analogous way (as in Eqns.
\eqref{e:nonkolmogorov}). Classical probabilities are a particular
case when the algebra is commutative
\cite{Redei-Summers2006,Gudder-StatisticalMethods}. In this way,
observables can be defined in theories more general than Hilbertian
QM.

\noindent One of the main differences between the axioms
\ref{e:kolmogorovian} and \ref{e:nonkolmogorov} is that the
$\sigma$-algebra in (\ref{e:kolmogorovian}) is boolean, while
$\mathcal{P}(\mathcal{H})$ is not. In this sense, the measures
defined by Eqns. \ref{e:nonkolmogorov} are called
\emph{non-kolmogorovian (or non-boolean) probability measures}.

\noindent One of the expressions of the fact that quantum and
classical probabilities are different, is that Eq. \eqref{e:SumRule}
is no longer valid in QM. Indeed, in QM it may happen that

\begin{equation}
s(A)+s(B)\leq s(A\vee B)
\end{equation}

\noindent von Neumann considered that Eq. \eqref{e:SumRule} was
crucial for the interpretation of $\mu(A)$ and $\mu(B)$ as relative
frequencies \cite{Redei99} in a frequentistic interpretation. But as
explained in \cite{Redei99,RedeiWhyVNDidNotLikedHilbert} one of the
main dissatisfactions of von Neumann was that Eq. \eqref{e:SumRule}
was not generally valid in the quantum case, making the
frequentistic interpretation untenable. This was one of the reasons
that led him to search for generalizations of the algebra of
projections in Hilbert space, and type II$_{1}$ factors were good
candidates for this objective \cite{RedeiWhyVNDidNotLikedHilbert}.

\subsection{A new approach to quantum
probabilities}\label{s:PaperAnnals}

In a recent work \cite{Holik-Saenz-Plastino} it was shown that the
approach to probability theory of R. T. Cox \cite{CoxLibro,CoxPaper}
can be applied to lattices more general than Boolean. And in
particular, that quantum probabilities and the generalized
probability theory can be obtained by applying a variant of this
method. In the rest of this work we explore possible interpretations
of this fact under the light of the problem posed by von Neumann
(Section \ref{s:QuotationLarge} of this work)\footnote{See also
\cite{Symmetry} for a very interesting but different perspective of
the R. T. Cox approach and quantum probabilities.}. We don't have
place here to introduce all the details (for which we refer to
\cite{Holik-Saenz-Plastino}) and just limit ourselves to describe
the general method:

\begin{itemize}
\item Our starting point is an orthomodular lattice
$\mathcal{L}$.

\item Next, we assume that $\mathcal{L}$ represents the propositional structure of a given system.

\item It is reasonable to assume\footnote{As a precondition of physical science, something which is
not necessarily true in any field of experience. A similar remark
holds for the existence of ---at least--- statistical regularities:
if such regularities are not present, mathematical description of
phenomena is untenable. We are not asserting that any phenomena
could be subsumed into this condition, but that it is a precondition
of mathematical physics.} that there is a definite state of affairs
determined by the preparation of the system. This preparation could
be natural or artificial, this is not relevant. But the system has
its own definite history as a constitutive feature of its own
actuality.

\item Define a function $s:\mathcal{L}\longrightarrow
\mathbb{R}$ such that $s(a)\geq 0$ $\forall a\in\mathcal{L}$ and it
is order preserving ($a\leq b \longrightarrow s(a)\leq s(b)$). This
function is intended to represent the degree of likelihood about
what would happen in the different (contextual) future situations.
But it is important to remark that this measure is a manifestation
of a structured actual state of affairs: its origin is ontological
and there are no hidden variables.
\end{itemize}

\noindent It can be shown that under the above rather general
assumptions, a probability theory can be developed
\cite{Holik-Saenz-Plastino} following a variant of R. T. Cox
approach \cite{CoxLibro,CoxPaper}. In other words, it is possible to
show that:

\begin{subequations}\label{e:GeneralDeducedProbability}
\begin{equation}\label{e:Ortogonal1}
s(\bigvee\{a_i\}_{i\in\mathbb{N}})=\sum_{i=1}^{\infty}s(a_i)
\end{equation}
\begin{equation}
s(\neg a)= 1-s(a)
\end{equation}
\begin{equation}
s(\mathbf{0})=0
\end{equation}
\end{subequations}

\noindent (where the $\{a_{i}\}$ in Eqn. \eqref{e:Ortogonal1} form a
denumerable and orthogonal family). Let us see an example of how the
Cox's machinery works. If $a,b\in\mathcal{L}$ and $a\bot b$, we have
that $a\wedge b=\mathbf{0}$. Next, it is reasonable to assume that
$s(a\vee b)$ can only be a function of $s(a)$ and $s(b)$. In this
way, $s(a\vee b)=f(s(a),s(b))$, with $f$ an unknown function to
determine. Due to associativity of ``$\vee$", $s((a\vee b)\vee
c)=s(a\vee(b\vee c))$ for any $a,b,c\in\mathcal{L}$. If $a$, $b$ and
$c$ are orthogonal, we will have $s((a\vee b)\vee
c)=f(f(s(a),s(b)),s(c))$ and $s(a\vee (b\vee
c))=f(s(a),f(s(b),s(c)))$. But then
$f(f(s(a),s(b)),s(c))=f(s(a),f(s(b),s(c)))$. Or put in a more simple
form, we are looking for a function $f$ such that

\begin{equation}\label{e:FunctionalEquation}
f(f(x,y),z)=f(x,f(y,z))
\end{equation}

\noindent But Eqn. \eqref{e:FunctionalEquation} is a functional
equation \cite{aczel-book} whose solution ---\emph{up to
rescaling}\footnote{For a discussion about the rescaling we refer to
\cite{CoxLibro}}--- is $f(x,y)=x+y$. In this way we arrive at
$s(a\vee b)=s(a)+s(b)$. Eqns. \eqref{e:GeneralDeducedProbability}
follow in a similar way \cite{Holik-Saenz-Plastino}.

\noindent As explained above, it is possible to show that the
probability theory defined by Eqns.
\ref{e:GeneralDeducedProbability} is non classical in the general
case. If $\mathcal{L}$ is not Boolean, it may happen that
$s((a\wedge\neg b)\vee(a\wedge b))=s(a\wedge\neg b)+s(a\wedge b)\leq
s(a)$, but any Kolmogorovian probability satisfies $s(a)=s(a\wedge
b)+s(a\wedge\neg b)$ \cite{Holik-Saenz-Plastino}.

\section{Probability theory as arising as the logical and geometrical structuration of phenomena}\label{s:GLP}

Maybe it is not just a coincidence that the discussion posed by von
Neumann on Logic, Probability and Geometry appeared in the
axiomatization of quantum theory. QM seems to pose a problem in the
interpretation of space-time, as is expressed, for example, in the
impossibility of defining trajectories for the particles. In this
way, a new kind of structure of experience underlies the quantum
mechanical description.

Space-time ---as considered by modern physics--- is not a naturally
given structure: it was a great achieving to develop geometry up to
the point to which it is possible to give the mathematical
description of reality provided for example, by classical mechanics
(CM) or general relativity (GR). The continuous description of
experience provided by Euclidean geometry, has as a precondition to
have definite logical objects: mathematical objects such as numbers,
geometrical figures, and all of this related to things of our
experience. Our experience in not a complete chaos and can be
structured in such a way. But we must never forget that the fact
that we can organize our experience in a space-time description is
just an assumption whose consistency is to be tested empirically.
General relativity, shows us that one can use a more elegant and
more powerfully predictive description of experience than the one
provided by the flat space-time of the Euclidean geometry of
classical physics. But the limits and success of these descriptions
are not granted in advance: they must be confronted with their
capability of defining a consistent experience.

But we are committed to the space time description in the following
sense: we need definite things and objective things to happen in
order to even \emph{speak} about an experiment. An example of this
is a pointer of an instrument yielding a value in a given outcome
set (which could be, for example, the set of real numbers, but could
also be more general, like the set formed by $\{+,-\}$). The fact
that an outcome set always forms a set and the events will be
represented by its subsets (forming a $\sigma$-algebra), ties us to
a very specific kind of logic (classical) and a very specific form
of spaciality (example, Euclidean geometry, or a curved space-time).
This is the real content of the observations of N. Bohr: the very
possibility of exerting experiments ties us to classical logic and a
set theoretical organization of experience. Space-time description
is just a particular case of this more general regulative logical
machinery.

But there is absolutely nothing granting us that this Boolean
description (thought necessary to exert experiments) will exhaust
the scenario in which phenomena appear. And this lies at the heart
of the existence of complementary (and incompatible) contexts in
quantum mechanics: in order to determine the state of the system, a
quantum tomography must be exerted, and thus, we are obliged to
study the system in different incompatible contexts. While the
structuration of experience in CM can be reduced to a boolean
algebra (and thus, to a set theoretical description provided by an
outcome set), in QM this is no longer possible. In classical
mechanics, the description of an object can be equated with its
space-time representation: form the point of view of classical
mechanics, the main goal is to describe continuous motion of
material bodies inside space. That is why motion (and change) can be
described as the solutions of deterministic differential equations.
And this feature is much more general than the usual description of
a particle moving through space under the action of forces. Any
quantity of interest taking continuous values, if it is classical,
will have associated a time derivative, and thus the description
reduces to the motion of a system in a phase space obeying
deterministic differential equations. Quite contrarily, quantum
mechanics is characterized by \emph{jumps}, by \emph{discontinuous}
and \emph{unpredictable} behavior. That is why the organization of
experience in QM comes endowed with a probabilistic description: it
is impossible to predict the future events with complete certainty,
and thus, the actual state of affairs \emph{is just a probability
distribution}.

In this way, QM fails to give a spatio-temporal description of
phenomena. In other words, \emph{QM shows us that the
spatio-temporal description is just a part (or perspective) of the
whole scenario of the organization of phenomena}; one of the most
important consequences of quantum mechanics is that space-time can
no longer considered as an exhaustive scenario in which physical
events take place. Quite on the contrary, experience can be
structured as a logic, and at the same time as a specific kind of
geometry, as von Neumann explained. Different models of event
structures represent different organizations of phenomena.

The results of \cite{Holik-Saenz-Plastino} show that once the
logic-algebraic properties of the structured experience are
determined, to great extent, the whole probability theory is
determined. In this way, we have a concrete step in the solution of
the problem posed by von Neumann. Experience is not complete chaos,
but on the contrary, it can be structured. This organization of
phenomena may have a definite \emph{logical form} (as is the case in
the CM or the QM descriptions), and this \emph{form} is expressed as
a \emph{geometry}.

But in QM this geometry must not be confused with the geometrical
background of space-time (Euclidean space or the curved background
of general relativity); space-time description is just an aspect or
perspective of a more general state of affairs. In the general case,
physical events can be organized as lattices\footnote{Or even more
generally, $\sigma$-orthocomplemented orthomodular posets
\cite{Gudder-StatisticalMethods}, which are not lattices in the
general case.} much more general than the boolean case and a similar
assertion holds.

Thus, once that the structure of experience is determined as a
Logic-Geometry, a probability calculus follows. If the description
is boolean, then probabilities will be Kolmogorovian, and
deterministic equations of motion can be used (in principle) to
govern the laws of motion. But if the logic is (ontologically
speaking) non-boolean, then the description will fail to be
deterministic because an ignorance interpretation of probabilities
will be untenable. Thus, deterministic equations of motion (as the
Schr\"{o}dinger equation) must be complemented with ``jumps" (as the
quantum jumps) and the concomitant \emph{processes} that they
trigger.

Now, a crucial question is in order: which kind of objects fit with
this notion of structured experience? In other words, which kind of
objects are compatible with the organization of events provided by
QM (or more general non-boolean lattices)? Our answer is that in the
QM description, objects appear as a partial aspect of a particular
description \footnote{There is no room in this small article to
discuss with full detail: this task will be completed elsewhere}. As
an example, think about a room full of objects. Each object (the
door, the walls, the chairs, a source, a photon counter, etc) has a
definite position and is situated in a definite relationship with
respect to the others. But which is the nature of the room itself?
The room itself, as it presents to us, comes into being as an
\emph{organized structure} of objects: \emph{everything is
correlated in some way}. To presuppose that the description of the
room can be reduced to the relative positions of objects in
Euclidean space (or in a more general spatio-temporal setting), is a
metaphysical assumption which is not necessarily valid for the
description of all phenomena\footnote{And of course, there are lots
of rooms which can be described using classical mechanics}. A
quantum system is just as real as the \emph{place} in which objects
are situated and structured, but it is not an object: \emph{it is
the organization of phenomena itself}. There is an actual state of
affairs in the room, which has its own history (state), and the
set-theoretical-spatio-temporal description as a collection of
objects in space is just an \emph{aspect} of it. And this structure
is logical form expressed as a particular geometry. A quantum setup
in a laboratory cannot be reduced to the classical description: this
is at the heart of the complementarity principle. In this way the QM
description manifests itself as the study of probability
distributions, which are of course, objective and (at least in
principle), experimentally controllable.

\end{document}